\documentclass[12pt]{article}
\usepackage[]{epsfig}
\usepackage[centertags]{amsmath}
\usepackage{amsfonts}
\usepackage{amssymb}
\usepackage[english]{babel}
\usepackage{amsmath, amsthm, slashed,url}
\usepackage{color}
\begin{document}

\title{\bf Analytic solution for Gauged Dirac-Weyl equation in $(2+1)$-dimensions }
\author{Juan Sebasti\'an Ardenghi, Alfredo Juan, Federico Escudero and Lucas Sourrouille
\\
{\normalsize\it  IFISUR, Departamento de F\'isica (UNS-CONICET) 
}\\ {\normalsize\it Avenida Alem 1253, Bah\'ia Blanca, Buenos Aires, Argentina}
\\
{\footnotesize  sourrou@df.uba.ar} } \maketitle

\abstract{A gauged Dirac-Weyl equation in (2+1)-dimension is considered. This equation has the particularity to describe 
the states of a graphene Dirac matter. In particular we are interested in matter interacting with a Chern-Simons gauge
fields. We show that exact self-dual solutions are admitted. These solutions are the same as those supported by nonrelativistic 
matter interacting with a Chern-Simons gauge field. }
 
\vspace{0.3cm}


\vspace{1cm}
\section{Introduction}

The two dimensional matter field interacting with gauge fields whose dynamics is governed by a Chern-Simons term support soliton
solutions\cite{Jk,Jk1,hor,hor1,hor2,hor3,hor4}. These models have the particularity to became auto-dual when the 
self-interactions are suitably 
chosen \cite{JW,JW1,JP,JP1}. When this occur the model presents particular mathematical and physics properties, such as the 
supersymmetric 
extension of the model \cite{LLW}, and the reduction of the motion equation to first order derivative equation \cite{Bogo}. The 
Chern-Simons gauge field inherits its dynamics from the matter fields to which it is coupled, so it may be either 
relativistic \cite{JW} or non-relativistic \cite{JP,JP1}. In addition the soliton solutions are of topological and 
non-topological 
nature \cite{JLW}.
\\
In the present Letter, we investigate Dirac-Weyl massless fermions under 
perpendicular magnetic field whose dynamics is dictated 
by a 
Chern-Simons gauge field.
In particular we show that this gauge theory admit soliton solutions which are analytic and coincide with the self-dual 
solutions supported by 
Schr\"{o}dinger-Chern-Simons model \cite{JP,JP1} defined by the Lagrangian 
density 
\begin{eqnarray}
\mathcal{L} = \frac{\kappa}{2}\epsilon^{\mu \nu \rho} A_\mu \partial_\nu A_\rho
+i\psi^* D_0 \psi - \frac{1}{2m} |D_i \psi|^2 + \frac{g}{2}|\psi|^4 
\label{jackiw-pi}
\end{eqnarray}
where, the first term is a Chern-Simons gauge field dynamics, which are coupled to nonrelativistic 
bosonic matter, represented  by the complex scalr field $\psi$. 

\section{The soliton solution}
\begin{figure}[tbp]
\centering\includegraphics[width=120mm,height=75mm]{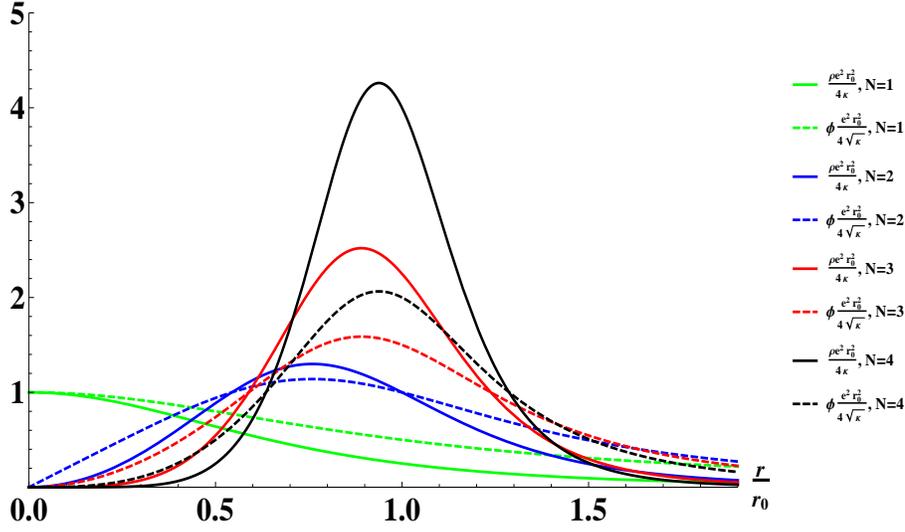}
\caption{Scalar field density $\rho $ and self-dual field $\phi$ as a function of $r/r_0$ for different values of 
$N$ and $\theta=0 $.}
\label{figure1}
\end{figure}

Let us start by considering a $(2+1)$-dimensional Dirac-Weyl-Chern-Simons model coupled to two-component spinors
\begin{equation}
\Psi=(\psi_a,\psi_b)^T
\label{}
\end{equation}
where $\psi_a$ and $\psi_b$
represent the envelope functions associated with the probability amplitudes. In addition, T denotes the transpose of the column 
vector.  
Then the action is governed by 
\begin{equation}
S= \int d^{3}x \Big( \frac{\kappa}{2}\epsilon^{\mu \nu \rho} A_\mu \partial_\nu A_\rho + \Psi^\dagger \sigma^0 D_0 
\Psi - \Psi^\dagger \sigma^i D_i \Psi \Big)
\label{Ac1}
\end{equation}
Here, the covariant derivative is defined as $D_{0}= -i\partial_{0} 
-eA_{0}$, $D_{i}= -i\partial_{i} +eA_{i}$ $(i =1,2)$, the metric tensor is  $g^{\mu \nu}=(-1,1,1)$ and 
$\epsilon^{\mu\nu\lambda}$ is the totally antisymmetric tensor such that $\epsilon^{012}=1$. Also, $\sigma^\mu$ $(\mu =0,1,2)$ 
are 2$\times$2 Pauli matrices, i.e.
\begin{eqnarray}
\sigma^0 =\left( \begin{array}{cc}
1 & 0 \\
0 & 1 \end{array} \right)
\,,
\;\;\;\;\;\
\sigma^1 =\left( \begin{array}{cc}
0 & 1 \\
1 & 0 \end{array} \right)
\,,
\;\;\;\;\;\
\sigma^2 =\left( \begin{array}{cc}
0 & -i \\
i & 0 \end{array} \right)
\end{eqnarray}
The Chern-Simons term of the action (\ref{Ac1}) may be developed integrating by parts,
\begin{equation}
S_{cs}=\frac{\kappa}{2}\int d^{3}x \Big( \epsilon^{\mu \nu \rho} A_\mu \partial_\nu A_\rho \Big)= 
\kappa \int d^{2}x \Big(A_0 F_{12} + A_2 \partial_0 A_1\Big)
\label{Ac2}
\end{equation}
On the plane the curl of a vector is a scalar, so that the magnetic field is $F_{12}= \partial_1 A_2 - \partial_2 A_1$. 
We may also develop the Dirac-Weyl term,
\begin{eqnarray}
S_{dw}&=& \int d^{3}x \Big( \Psi^\dagger \sigma^0 D_0 
\Psi - \Psi^\dagger \sigma^i D_i \Psi \Big)
\nonumber \\
&=&\int d^{3}x \Big( -i\psi_a^\dagger \partial_0 \psi_a -eA_0 
\psi_a^\dagger\psi_a -i\psi_b^\dagger \partial_0 
\psi_b -eA_0 \psi_b^\dagger\psi_b -[
-i\psi_a^\dagger\partial_1\psi_b -\psi_a^\dagger \partial_2 \psi_b 
\nonumber \\
&&+ eA_1 \psi_a^\dagger \psi_b
-ie A_2 \psi_a^\dagger \psi_b - i\psi_b^\dagger \partial_1 \psi_a + \psi_b^\dagger \partial_2 \psi_a + eA_1\psi_b^\dagger \psi_a 
+ ieA_2 \psi_b^\dagger \psi_a] \Big)
\label{Ac3}
\end{eqnarray}
Then, the corresponding field equations for the action (\ref{Ac1}) are
\begin{eqnarray}
&-&i\partial_0 \psi_a -eA_0 \psi_a - [-i\partial_1 \psi_b - \partial_2 \psi_b + eA_1 \psi_b -ieA_2 \psi_b] 
\nonumber \\
&=& D_0 
\psi_a -[D_1 \psi_b -iD_2 \psi_b] =0
\label{eqm1}
\end{eqnarray}
\begin{eqnarray}
&-&i\partial_0 \psi_b -eA_0 \psi_b - [-i\partial_1 \psi_a + \partial_2 \psi_a + eA_1 \psi_a -ieA_2 \psi_a]
\nonumber \\
&=& D_0 
\psi_b - [D_1 \psi_a +iD_2 \psi_a] =0
\label{eqm2}
\end{eqnarray}
\begin{eqnarray}
&&\kappa F_{12} -e[\psi^\dagger_a \psi_a + \psi^\dagger_b \psi_b] =0
\label{eqm3}
\end{eqnarray}
\begin{eqnarray}
&&\kappa (-\partial_0 A_2 + \partial_2 A_0) - e[\psi^\dagger_a \psi_b + \psi^\dagger_b \psi_a] =0
\label{eqm4}
\end{eqnarray}
\begin{eqnarray}
&&\kappa (\partial_0 A_1 - \partial_1 A_0) - e[-i\psi^\dagger_a \psi_b + i\psi^\dagger_b \psi_a] =0
\label{eqm4.1}
\end{eqnarray}
The equations (\ref{eqm1}) and (\ref{eqm2}) may be expressed in a compact form as
\begin{eqnarray}
\sigma^0 D_0 \Psi - \sigma^i D_i \Psi =0
\label{eqm5}
\end{eqnarray}
which is the massless Dirac-Weyl equation in (2+1)-dimensions. This equation is gauge invariant since a gauge transformation of 
the potentials,
\begin{eqnarray}
A_i \rightarrow A_i - \frac{1}{e} \partial_i \omega \,,
\;\;\;\;\;\
A_0 \rightarrow A_0 + \frac{1}{e} \partial_0 \omega
\label{ge}
\end{eqnarray}
accompanied by a transformation of the spinor 
\begin{eqnarray}
\Psi \rightarrow e^{i \omega} \Psi
\label{ge1}
\end{eqnarray}
leaves the equation (\ref{eqm5}) unchanged.
\\
The gauge field satisfied its dynamical equations, which are dictated by the formulas (\ref{eqm3}), (\ref{eqm4}) and 
(\ref{eqm4.1}). 
These are the Chern-Simons field equations coupled to matter field by $j^0= e[\psi^\dagger_a \psi_a + \psi^\dagger_b \psi_b]$ 
and $j^i = e\psi^\dagger \sigma^i \psi$, which are the conserved 
currents associated to gauge symmetry (\ref{ge1}). So that,
\begin{eqnarray}
\partial_\mu j^\mu = \partial_0 j^0 + \partial_i j^i =0
\label{}
\end{eqnarray}
In particular, the field equations (\ref{eqm3}), (\ref{eqm4}) and (\ref{eqm4.1}) may be reduced to a single equation 
\begin{eqnarray}
\frac{\kappa}{2} \epsilon^{\nu \alpha \beta} F_{\alpha \beta}= j^\nu
\label{cs}
\end{eqnarray}
Thus, the equation (\ref{eqm3}) is the time component of this equation 
\begin{eqnarray}
\kappa F_{12} = j^0
\label{gas}
\end{eqnarray}
Then, integrating over the entire plane, we obtain the important consequence that any field configuration with charge $Q =\int 
d^2 x j^0$ also carries magnetic flux $\Phi = \int B d^2 x$ 
\cite{Echarge,Echarge1,Echarge2}:
\begin{eqnarray}
\Phi =\frac{1}{\kappa} Q
\end{eqnarray}
In addition, the equations (\ref{eqm4}) and (\ref{eqm4.1}) are 
the spatial components of (\ref{cs}),
\begin{eqnarray}
j^i = \epsilon^{i j} \kappa E_j
\end{eqnarray}
\\
In this note we will show that the system (\ref{Ac1}) admit a static soliton solution carrying magnetic flux and electric 
charge. In order to show this we consider the stationary points of the action which for the static field configuration 
reads
\begin{eqnarray}
S= \int d^{2}x \Big(\kappa A_0 F_{12} -eA_0 
(\psi_a^\dagger\psi_a + \psi_b^\dagger\psi_b) -[
-i\psi_a^\dagger\partial_1\psi_b -\psi_a^\dagger \partial_2 \psi_b
+ eA_1 \psi_a^\dagger \psi_b
\nonumber \\
-ie A_2 \psi_a^\dagger \psi_b - i\psi_b^\dagger \partial_1 \psi_a + \psi_b^\dagger \partial_2 \psi_a + eA_1\psi_b^\dagger \psi_a 
+ ieA_2 \psi_b^\dagger \psi_a]\Big)
\label{Ace}
\end{eqnarray}
In view of Gauss law constraint (\ref{gas}), the action may be rewritten as
\begin{eqnarray}
S= -\int d^{2}x \Big(
-i\psi_a^\dagger\partial_1\psi_b -\psi_a^\dagger \partial_2 \psi_b
+ eA_1 \psi_a^\dagger \psi_b -ie A_2 \psi_a^\dagger \psi_b 
\nonumber \\
- i\psi_b^\dagger \partial_1 \psi_a + \psi_b^\dagger \partial_2 \psi_a + eA_1\psi_b^\dagger \psi_a 
+ ieA_2 \psi_b^\dagger \psi_a \Big)
\nonumber \\
= -\int d^{2}x \Big(\psi_a^\dagger [D_1 \psi_b -iD_2 \psi_b] + \psi_b^\dagger [D_1 \psi_a + iD_2 \psi_a] \Big)
\label{Ace1}
\end{eqnarray}
To proceed, we can use the static version of equations (\ref{eqm1}) and (\ref{eqm2}), i.e.,
\begin{eqnarray}
\psi_a = \frac{-1}{eA_0} [D_1 \psi_b -iD_2 \psi_b] 
\label{}
\end{eqnarray}
\begin{eqnarray}
\psi_b = \frac{-1}{eA_0} [D_1 \psi_a +iD_2 \psi_a]
\label{}
\end{eqnarray}
So, the equation (\ref{Ace1}) reads
\begin{eqnarray}
S &=& \int d^{2}x \frac{1}{eA_0}\Big(  [D_1 \psi_b -iD_2 \psi_b]^\dagger [D_1 \psi_b -iD_2 \psi_b] + 
[D_1 \psi_a +iD_2 \psi_a]^\dagger [D_1 \psi_a + iD_2 \psi_a] \Big)
\nonumber \\
&=& \int d^{2}x \frac{1}{eA_0}\Big( |D_1 \psi_b -iD_2 \psi_b|^2 + |D_1 \psi_a + iD_2 \psi_a|^2 \Big)
\label{Ace2}
\end{eqnarray}
Since, $A_0$ is a Lagrange multiplier, which does not play any role in the search of static solution, we can choose, 
without loss of 
generality, $A_0>0$. Thus, the action (\ref{Ace2}) is non-negative and bounded below by zero. This lower bound 
is saturated by solutions to the first-order self-duality equations 
\begin{eqnarray}
(D_1  -iD_2) \psi_b = 0
\nonumber \\[3mm]
(D_1  + iD_2) \psi_a =0
\end{eqnarray}
Together with the Gauss law (\ref{gas}) these two equations compose the set of the field equations whose solutions minimize the 
static action (\ref{Ace}). In particular, an interesting situation emerge when one of the spinor component is set to zero. In 
that case, the set of equations reduces to the self-duality equations of the Schr\"{o}dinger-Chern-Simons model present in 
Ref.\cite{JP,JP1}, 
\begin{eqnarray}
(D_1  + iD_2) \psi_a =0
\nonumber \\[3mm]
\kappa F_{12} = e \psi^\dagger_a \psi_a 
\label{du1}
\end{eqnarray}
with $\psi_b =0$ or 
\begin{eqnarray}
(D_1  -iD_2) \psi_b = 0
\nonumber \\[3mm]
\kappa F_{12} = e \psi^\dagger_b \psi_b
\label{du2}
\end{eqnarray}
with $\psi_a =0$.
Both, (\ref{du1}) and (\ref{du2}) may be solved analytically. We can take the set 
(\ref{du1}). To solve these equations is usual to decompose the scalar field $\psi_a$ into its phase and magnitude:
\begin{eqnarray}
\psi_a = \rho^{\frac{1}{2}} e^{i\alpha}
\label{}
\end{eqnarray}
where, $\rho = \psi^\dagger_a \psi_a$. Then, multiplying the first of the self-duality equations (\ref{du1}) by $i\psi^\dagger_a$ 
and its complex conjugate by $-i\psi_a$ we arrive to
\begin{eqnarray}
i\psi^\dagger_a(D_1  + iD_2) \psi_a -i\psi_a[(D_1  + iD_2) \psi_a]^\dagger = 2ieA_2|\psi_a|^2 -2|\psi_a|^2 i\partial_2 \alpha + 
i\partial_1 |\psi_a|^2 =0
\end{eqnarray}
\begin{eqnarray}
i\psi^\dagger_a(D_1  + iD_2) \psi_a +i\psi_a[(D_1  + iD_2) \psi_a]^\dagger = 2ieA_1|\psi_a|^2 -2|\psi_a|^2 i\partial_1 \alpha - 
i\partial_2 |\psi_a|^2 =0
\end{eqnarray}
This, determines the gauge field
\begin{eqnarray}
A_i = \frac{1}{2e \rho}\Big( \epsilon^{ij} \partial_j (\log \rho)- 2 \partial_i \alpha \Big)
\label{a11}
\end{eqnarray}
everywhere away from the zeros of the scalar field. Thus, using (\ref{a11}) the second self-duality equation of (\ref{du1}) 
reduces to a nonlinear elliptic equation for the scalar field density $\rho$,
\begin{eqnarray}
\nabla^2 \log \rho = -\frac{2 e^2}{\kappa} \rho
\label{}
\end{eqnarray}
We can proceed in similar way and take the set (\ref{du2}). Then, we arrive to
\begin{eqnarray}
\nabla^2 \log \rho = \frac{2 e^2}{\kappa} \rho
\label{}
\end{eqnarray}
These are elliptic equations, known as the Liouville equations and are exactly solvable,
\begin{eqnarray}
\rho = \frac{\kappa}{e^2} \nabla^2 \log \Big(1 + |f|^2\Big)
\label{}
\end{eqnarray}
where $f=f(z)$ is a holomorphic function of $z= x_1 +ix_2$.
General radially symmetric solutions may be obtained by taking $f(z)=\Big(\frac{z_0}{z}\Big)^N$. 
Then, we have 
\begin{eqnarray}
\rho = \frac{4 \kappa N^2}{e^2 r_0^2} \frac{ \Big(\frac{r}{r_0}\Big)^{2(N-1)}}{\Big(1+ (\frac{r}{r_0})^{2N} \Big)^2}
\label{}
\end{eqnarray}
This vanish as $r\to\infty$ and is nonsingular at the origin for $|N|\geq 0$ but for  $|N|> 0$, the vector 
potential behaves as $A_i (r) \sim - \partial_i \alpha - 2(N-1)\epsilon_{ij}\frac{x^j}{r^2}$, which indicates that it has a 
singular contribution at $r=0$. The singularity may be avoided if we choose the phase of $\phi$ to be $\alpha = \theta(N-1)$. 
Then,
the self-dual $\phi$ field is (see figure \ref{figure1})
\begin{eqnarray}
\phi = \frac{\sqrt{\kappa}2 N }{e r_0} \Big(\frac{(\frac{r}{r_0})^{N-1}}{1+ (\frac{r}{r_0})^{2N}} \Big) e^{-
i(N-1)\theta}
\label{sol}
\end{eqnarray}
Requiring that $\phi$ be single-valued we find that $N$ must be an integer, and for $\rho$ to decay at infinity
we require that $N$ be positive.
\\
To conclude it is interesting to comment that the solution (\ref{sol}) is the same as the soliton solution discussed in 
Ref.\cite{JP,JP1}. In fact, as we mentioned, the self-duality equations (\ref{du1}) and (\ref{du2}) coincide with  the 
self-duality equations of the Schr\"{o}dinger-Chern-Simons model present in Ref.\cite{JP,JP1}. The reason for this lies in the 
fact that the static Hamiltonian associated to the model (\ref{jackiw-pi}) is 
\begin{eqnarray}
H = \int \,\, d^2 x & \Big(  \frac{1}{2m} |D_i \psi|^2 - \frac{g}{2}|\psi|^4 \Big)
\label{EJP}
\end{eqnarray}
The static solutions, which are the stationary points of the Hamiltonian, may be found in view of the Chern-Simons Gauss law 
$B=\frac{e}{\kappa}\rho$ and 
the identity
\begin{eqnarray}
|D_i \psi|^2 = |( D_1 \pm iD_2)\psi|^2 \mp eB|\psi|^2 \pm m\epsilon^{ij} \partial_i J_j
\label{iden}
\end{eqnarray}

Then, (\ref{EJP}) reads as
\begin{eqnarray}
H = \int \,\, d^2 x & \Big(  \frac{1}{2m} |( D_1 \pm iD_2)\psi|^2  \pm \frac{\epsilon^{ij}}{2} \partial_i J_j 
+ [-\frac{g}{2} \mp \frac{e^2}{2m\kappa}]|\psi|^4 \Big)
\label{EJP1}
\end{eqnarray}
Here, the second term in (\ref{EJP1}) is a surface term. To see this, we can apply the Stokes' theorem, then we have,
\begin{eqnarray}
\int \,\,d^2 x \epsilon^{ij} \partial_i J_j = \oint_{|x|=\infty} \,\, J_j dx^j 
\label{St}
\end{eqnarray}
where, $J_j = -\frac{i}{2m}\Big(\phi^* D_j \phi - (D_j \phi)^* \phi \Big)$. The requirement that the energy be finite states 
that the covariant derivative must vanish asymptotically. This fixes the behavior of the field at infinity. In the case of a 
nontopological theory such as the Jackiw-Pi model \cite{JP,JP1}, this implies the following boundary condition,
\begin{eqnarray}
\lim_{r \to \infty} \phi(x) = 0
\label{b1}
\end{eqnarray}
whereas the gauge field, at infinity, is a pure gauge. Hence, $J_j \to 0$ as $x \to \infty$. 
Thus, with the self-dual coupling
\begin{eqnarray}
g= \mp \frac{e^2}{m\kappa}
\end{eqnarray}
and sufficiently well behaved fields so that the integral over all space of $\epsilon^{ij} \partial_i J_j$ vanishes, the energy 
becomes
\begin{eqnarray}
E = \int \,\, d^2 x &  \frac{1}{2m} |(D_1 \pm iD_2) \psi|^2
\label{ef}
\end{eqnarray}
If we compare this expression with (\ref{Ace2}) we note that they are very similar. So, the fields that minimize the 
energy (\ref{ef}) are the same as minimize (\ref{Ace2}) and therefore obey the equations (\ref{du1}) and (\ref{du2}). Thus, 
the 
identity (\ref{iden}) plays and important role in order to connect the Dirac model with a non-relativistic model.    
\\
In summary, we show that the Dirac-Weyl field interacting with gauge fields governed 
by Chern-Simons dynamics, support analytic static self-dual solutions. 
The solutions that we found are the same as the solution supported by nonrelativistic matter interacting 
with a Chern-Simons gauge fields. In addition, it well know \cite{Jk1}-\cite{Jk9} that in 
the low energy electronic excitations of graphene, an expansion around any of the two fermi points gives an effective Hamiltonian 
linear in momentum which reduces to the massless Dirac equation in two dimensions derived from the Hamiltonian,
\begin{equation}
H = \upsilon_F \int d^{2}x \Big( \Psi^\dagger \sigma^i \partial_i \Psi \Big)
\label{}
\end{equation}
where $\upsilon_F = 8 \times 10^5 m/sec$ is the Fermi velocity. This Hamiltonian is associated to a static field configuration 
and therefore it can be derived from a more general Hamiltonian for time dependent fields
\begin{equation}
H = \upsilon_F \int d^{3}x \Big(  - \Psi^\dagger \sigma^0 D_0 
\Psi + \Psi^\dagger \sigma^i D_i \Psi \Big)
\label{}
\end{equation}
which is the Hamiltonian for the description of a graphene layer in presence of electric and magnetic fields (see for instance 
\cite{J1,J2,J2.1,J2.2,J2.3,J2.4,J2.5,J3}).
\\
On the other hand, Chern-Simons term becomes important in the description of fractional quantum
Hall effect (FQHE) in graphene \cite{J3.1,J3.2,J3.3,J3.4,J3.5,J3.6,J3.7,J3.8}. A way to understand the nature
of these states is provided by the composite Fermion(CF) theory \cite{J4} in which the state of the system is described in terms
of CF quasiparticles which correspond to electrons bound to an even number (2k) of vortices of flux quantum $\Phi_0 = 
\frac{hc}{e}$. Such a flux attachment can also be understood by carrying out Chern-Simon (CS) transformation on the 
electron field operators, which leads to the introduction of a topological CS vector potential a resulting in a CS magnetic 
field, which is proportional to the electron density $j^0= e[\psi^\dagger_a \psi_a + \psi^\dagger_b \psi_b]$ \cite{J5,J6}. In 
other words, the dynamics of the magnetic field is dictated by the Chern-Simons Gauss law (\ref{gas}). Thus, the 
Chern-Simons term is important because allows us to introduce a general flux tied to the electrons, and then it has its own 
dynamics.
In particular, many works have been done in the study of graphene Dirac 
electrons 
interacting with an external magnetic field \cite{6}-\cite{15}. In general numerical computation is required and some simple 
cases for an electron in the presence magnetic field are solve analytically \cite{13,15}. In this direction, we think that our 
result may be 
important because 
constitutes an exact solution for the description of graphene Dirac electrons in a magnetic field with its own gauge 
dynamics dictated by a Chern-Simons term.

\vspace{0.6cm}

{\bf Acknowledgements}
\\

This paper was partially supported by grants of CONICET (Argentina National Research Council) and Universidad Nacional del Sur 
(UNS)and by ANPCyT through PICT 1770, and PIP-CONICET Nos. 114-200901-00272 and 114-200901-00068 research grants, as well as by 
SGCyT-UNS., A. J., J. S. A. and L. S. are members of CONICET.

\end{document}